\documentclass{ifacconf}
\usepackage{amsmath}
\usepackage{amssymb}
\usepackage{graphicx}
\usepackage{lipsum}
\usepackage{graphicx}
\usepackage{natbib}
\usepackage{tikz}
\usepackage{array}
\usepackage{booktabs,adjustbox}
\usepackage{kantlipsum}

\newenvironment{definition}[1][Definition]{\begin{trivlist}
\item[\hskip \labelsep {\bfseries #1}]}{\end{trivlist}}

\newcommand{\finalcells}[2]{%
  \begingroup\sbox0{\begin{minipage}{3cm}\raggedright#1\end{minipage}}%
  \sbox2{\begin{minipage}{3cm}\raggedright#2\end{minipage}}%
  \xdef\finalheight{\the\dimexpr\ht0+\dp0+\smallskipamount\relax}%
  \xdef\finalheightB{\the\dimexpr\ht2+\dp2+\smallskipamount\relax}%
  \ifdim\finalheightB>\finalheight
    \global\let\finalheight\finalheightB
  \fi\endgroup
  \begin{minipage}[t][\finalheight][t]{3cm}\raggedright#1\end{minipage}&
  \begin{minipage}[t][\finalheight][t]{3cm}\raggedright#2\end{minipage}}
\begin{document}
\begin{frontmatter}

\title{Geometric Controllability of The Purcell's Swimmer and its Symmetrized Cousin}

\author[First]{Sudin Kadam}
\author[First]{Ravi N. Banavar}

\address[First]{Systems and Control Engineering Department, Indian Institute of Technology Bombay, Mumbai, India 400076
(e-mail: sudin@sc.iitb.ac.in, banavar@iitb.ac.in).}

\begin{abstract}
We analyse weak and strong controllability notions for the locomotion of the 3-link Purcell's swimmer, the simplest possible swimmer at low Reynolds number from a geometric framework. After revisiting a purely kinematic form of the equations, we apply an extension of Chow's theorem to analyze controllability in the strong and weak sense. Further, the connection form for the symmetric version of the Purcells' swimmer is derived, based on which, the controllability analysis utilizing the Abelian nature of the structure group is presented. The novelty in our approach is the usage of geometry and the principal fiber bundle structure of the configuration manifold of the system to arrive at strong and weak controllability notions.
\end{abstract}

\begin{keyword}
Purcell's swimmer, Locomotion, Principal Fiber Bundle, Nonlinear Controllability.
\end{keyword}

\end{frontmatter}
%===============================================================================
\section{Introduction} Swimming at micro scales is a topic of growing interest. A vast majority of living organisms are found to perform motion at microscopic scales. There has been a lot of research and a growing interest in exploring new and efficient ways to generate propulsion at these scales, see [\cite{becker2003self}], [\cite{dreyfus2005microscopic}]. A better understanding of the mechanism of swimming can lead to many useful applications in  several fields such as medicine, micro-machining or micro and nano technology. Microbial motion occurs in a fluid medium with very low Reynolds number, which is the ratio of the inertial to viscous forces acting on the swimmer's body.  The Reynold's number in such regimes is of the order of $10^{-4}$ [\cite{najafi2004simple}]. The environmental interactions experienced by such microorganisms are essentially different from those experienced by larger animals, which have prominent inertial effects. On the contrary, the viscous forces strongly dominate the motion at low Reynold's number.

E. M. Purcell, in his lecture on Life at Low Reynolds Number [\cite{purcell1977life}], presented a three-link swimmer that can propel itself at low Reynolds numbers. This swimmer can be considered as a simplified flagellum made of three slender rods articulated at two hinges. This proposition gave rise to a lot of research in modelling, control, optimal gait design etc. of this swimmer, see [\cite{tam2007optimal}], [\cite{passov2012dynamics}], [\cite{burton2013dynamics}], [\cite{avron2008geometric}], [\cite{melli2006motion}], and the references therein. 

[\cite{bloch2003nonholonomic}], [\cite{holm2009geometric}], [\cite{ostrowski1998geometric}] indicate that geometric mechanics and control theory play a crucial role in the analysis of robotic and animal locomotion . For a large class of locomotion systems, including underwater vehicles, fishlike swimming, flapping winged vehicles, spacecraft with rotors and wheeled or legged robots, it is possible to model the motion using the mathematical structure of a connection on a principal bundle, see  [\cite{cabrera2008base}], [\cite{david2012proportional}], [\cite{ostrowski1998geometric}].

Although it seems relevant to apply geometric mechanics and control theoretic concepts to the Purcell's swimmer, the literature shows few geometric-oriented approaches. One of the recent works on the Purcell's swimmer by [\cite{hatton2013geometric}] studies the problem in such a geometric framework by introducing the connection form for modelling it as a purely kinematic system. We shall base our work on this model for the basic Purcell's swimmer, and will proceed to controllability analysis in the geometric framework. Furthermore, we derive the connection form followed by the controllability analysis for the symmetrized version of the swimmer, also referred to as the symmetrized cousin of the Purcell's swimmer in [\cite{avron2008geometric}].

\subsection{Contribution:}
The main contributions of this paper consist of highlighting the particular geometry of a trivial principal fiber bundle that the configuration space of this swimmer follows. [\cite{giraldi2013controllability}] analyzes the controlability of the swimmer using Chow's theorem by modelling it as a control affine, driftless, non-linear system. In our work, the recognition of the geometric structure is utilized to bring the controllability results in the light of strong and weak notions, which gives a more complete description to controllability results for a locomotion problem. Moreover, the derivation of the connection form for the symmetrized version of the Purcell's swimmer, followed by identification of its structure group's Abelian nature to generate the local controllability results are the novel contributions this paper achieves.

\subsection{Organization of paper:}
In the next section we show the geometry that typical locomotion systems have, followed by the example of the Purcell's swimmer. In section 3, we review the kinematic model of this swimmer and define the ideas of strong and weak controllability. In section $4$, the expression for the connection form for the symmetrized version of the Purcell's swimmer is derived. We then identify $3$ sets on it's base space at which system shows different controllability properties.

\section{Geometry of configuration space of the Purcell's swimmer}
\begin{figure}[h!]
\centering
\includegraphics[scale=0.37]{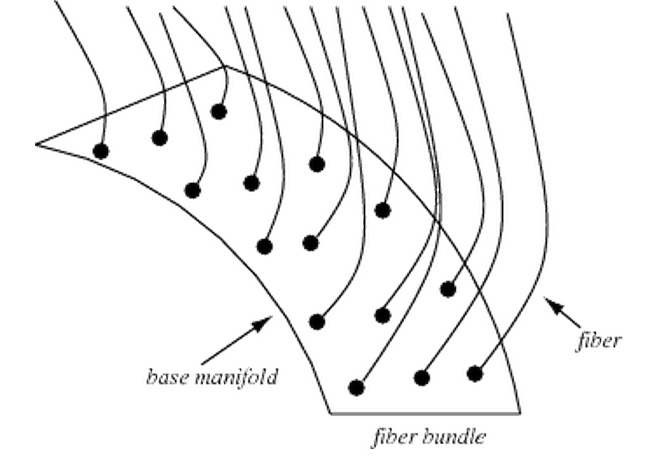} 
\caption{Fiber Bundle [\cite{weisstein}]}
\label{fiber_bundle}
\end{figure}

While studying problem of locomotion using internal shape change, the geometry of the configuration space, which is generally a differential manifold requires attention for an elegant and insightful solutions. The configuration space is written as the product of two manifolds, either locally or globally. One part, the base manifold $M$, describes the configuration of the internal shape variables of the mechanism. The other part depicts the macro-position of the locomoting body, a Lie group $G$, representing displacement of the body coordinate frame with respect to the reference frame. The total configuration space of the robot $Q$ is defined by both $G$ and $M$. Such systems follow the topology of a trivial principal fiber bundle, see [\cite{kobayashi1963foundations}]. Figure \ref{fiber_bundle} shows a schematic of a fiber bundle. With such a separation of the configruation space, locomotion is readily seen as the means by which changes in shape affect the macro position. We refer to [\cite{bloch1996nonholonomic}], [\cite{kelly1995geometric}] for more on the topology of locomoting systems. 

\begin{figure}[h!]
\centering
\includegraphics[scale=0.4]{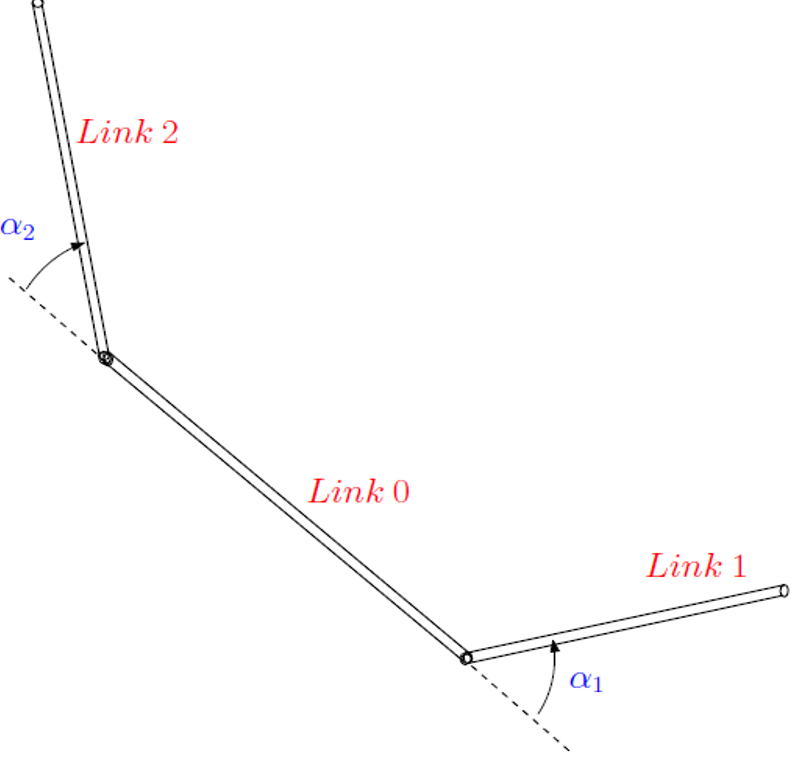}
\caption{Purcell's swimmer}
\label{original_purcell01}
\end{figure}

The Purcell's swimmer is a 3 link mechanism moving in a fluid with low Reynold's number. Each link of the swimmer is modelled as a rigid slender body of length $2L$. The original form of Purcell's swimmer, shown in fig. \ref{original_purcell01} has three links always in a common plane, the outer 2 links are actuated through respective rotary joints with base link. It is to be noted that we follow the notations similar to those in [\cite{hatton2013geometric}].

We represent the orientation of the outer links with respect to the base link through shape variables. The position of the three-link system in Fig. \ref{original_purcell01} is defined by location of the midpoint of the base link and its orientation. This is represented by $g$, which belongs to the Special Eucledean group $SE(2)$, parametrized by $(x,\: y,\: \theta)$. The shape space of mechanism is parametrized by the two joint angles $(\alpha_1, \alpha_2) \in \mathbb{S}^1\times\mathbb{S}^1$. Hence, the configuration space is 
\begin{equation}
Q\:=\:SE(2) \times\mathbb{S}^1\times\mathbb{S}^1
\end{equation}

\begin{definition}:
\textit{For $Q$ a configuration manifold and $G$ a Lie group, a trivial principal fiber bundle with base $M$ and structure group $G$ is a manifold $Q = M \times G$ with a free left action of $G$ on $Q$ given by left translation in the group variable: $\phi_h(x,g) = (x,hg)$ for $x \in M$ and $g \in G$.} [\cite{kelly1995geometric}]
\end{definition}

The structure group in our case is $G = SE(2)$, which is a matrix Lie group. The shape space $M$ is parametrized by $(\alpha_1, \alpha_2)$. Since all the points $q \in Q$ are represented by $(\alpha_1, \alpha_2, g)$ with $g \in G$,  $Q$ has global product structure of the form $M \times G$. Moreover, $SE(2)$ acts via left action as a matrix multiplication, and has a single identity element, which is a $3 \times 3$ identity matrix. Hence left action of group, defined by $\Phi_h : (x,g) \in Q \longrightarrow (x,hg)$ is free, for $x \in M$  and $\: h,g \in G$.

Thus, according to the definition, the configuration space of the basic Purcell's swimmer satisfies the trivial principal fiber bundle structure.

\section{Basic Purcell's swimmer}

\subsection{Reconstruction equation and local connection form}
In this section we briefly revisit the model of the Purcell's swimmer explained in [\cite{hatton2013geometric}]. It uses Resistive Force Theory, according to which hydrodynamic forces exerted on the swimmer can be approximated with local drag forces, which depend linearly on the velocity of each point [\cite{friedrich2010high}]. The resulting model is in a purely kinematic form, equation ($\ref{pure_kinematic}$). See [\cite{shammas2007geometric}] for details on purely kinematic systems. 
\begin{equation}\label{pure_kinematic}
\xi = -A(x)\dot{x}
\end{equation}
where $A(x)$ is the local connection form defined at each $x \in M$. For the Purcell's swimmer local connection form $A(x)$ is a $3 \times 3$ matrix which appears in the form of $\omega_1^{-1}\omega_2$. The matrices $\omega_1$ and $\omega_2$ are of size $3 \times 3$ and $3 \times 2$, respectively, and they depend on the lengths of the limbs, viscous drag coefficient $k$ and the shape of mechanism $(\alpha_1, \alpha_2)$. We refer to the [\cite{hatton2013geometric}] for their explicit form. We recall that for the shape manifold $M$, its tangent space at a point $x \in M$ is denoted by $T_xM$, and the shape velocity $\dot{x} = (\dot{\alpha}_1,\dot{\alpha}_2) \in T_xM$. The local connection form is thus defined as $A(x) : T_xM \longrightarrow \xi$. In our example the Lie group is the Special Eucledean group $SE(2)$, and $\xi=[\xi_x, \xi_y, \xi_{\theta}]^T$ belongs to its tangent space at the identity, with $\xi_x, \xi_y$ being the translational velocity of the base link and $\xi_{\theta}$ is its rotational component. The connection form and the other notions mentioned here have roots in geometric mechanics, see [\cite{bloch2003nonholonomic}], [\cite{holm2009geometric}] for details.

\subsection{Controllabitlity analysis}
We now proceed to analyze the controllability of the Purcell's swimmer in geometric setting. Since control in our case is the shape velocity itself, (\ref{pure_kinematic}) can be written in driftless control affine form as -

\begin{equation}\label{pure_kinematic2}
\begin{bmatrix}
\dot{x} \\
\xi
\end{bmatrix} = \begin{bmatrix}
I \\
A(x)
\end{bmatrix} u
\end{equation}
where, the negative sign in equation (\ref{pure_kinematic}) is absorbed in $A(x)$, $I$ is a $2 \times 2$ identity matrix and $u = [\dot{\alpha}_1, \dot{\alpha}_2]^T \in T_xM$ is the control input.

\subsection{Strong and weak controllability}
Since our configuration space is naturally split into shape and structure group, we write a point in configuration space as $q = (x,g) \in M \times G = Q$. We recall that for a curve $x(t) \in M$, the horizontal lift $x^*(t) \in Q$ is a curve which projects to $x(t)$ under the projection map defining principal fiber bundle and components of its tangent vectors $\dot{x}^*(t) \in T_qQ$ satisfy the reconstruction equation (\ref{pure_kinematic}). We refer to [\cite{kelly1995geometric}] for more details on this and also for the following 2 controllability notions.

\begin{itemize}
\item \textit{A locomotion system is said to be strongly controllable if, for any initial $q_0=(x_0,g_0)$ and final $q_f=(x_f,g_f)$, there exists a time $T > 0$ and a curve passing through $q_0$ satisfying $x^*(0) = q_0$ and $x^*(T) = q_f$.}

\item \textit{A locomotion system is said to be weakly controllable if, for any initial position $g_0 \in G$, and final position $g_f \in G$, and initial shape $x_0 \in M$, there exists a time $T > 0$ and a base space curve $x(t)$ satisfying $x(0) = x_0$ such that the horizontal lift of $x(t)$ passing through $(x_0, g_0)$  satisfies $x^*(0) = q_0$ and $x^*(T) = (x(T), g_f)$.}
\end{itemize}

As mentioned in the introduction,  [\cite{giraldi2013controllability}] analyzes controllability of the Purcell's swimmer by applying Chow's theorem (see [\cite{bullo2004geometric}] to the system in equation [\ref{pure_kinematic2}] by treating it as a driftless, control-affine system. It does so by checking span of the space formed by the successive Lie brackets of the vector fields corresponding to the two control inputs $\dot{\alpha}_1$ and $\dot{\alpha}_2$. In our approach, we rather utilize the principal fiber bundle structure, which gives rise to these strong and weak controllability notions. These define controllability ideas in more detail, and are of practical relevance since many times just reaching the desired group component without strict requirement on shape of the system is sufficient. We define following vector spaces, referred from [\cite{kelly1995geometric}].

\begin{align*}
\mathfrak{h}_1\:&=\: span \{A(X) : X \in T_x M\},\\
\mathfrak{h}_2\:&=\: span \{DA(X,Y) : X, Y \in T_x M\}, \\
\mathfrak{h}_3\:&=\: span \{L_Z DA(X,Y) - [A(Z),DA(X,Y)], \\ & \qquad \qquad \: [DA(X,Y),DA(W,Z)] : W, X, Y, Z \in TM\} \\
\vdots \\
\mathfrak{h}_k\:&=\: span \{L_X \xi - [A(X),\xi],[\eta,\xi] :
X \in T_x M, \xi \in \mathfrak{h}_{k-1}, \\
& \:\: \qquad\qquad \eta \in \mathfrak{h}_2 \:\oplus\: \cdots \:\oplus\:\mathfrak{h}_{k-1}\}
\end{align*}

Then a system defined on a trivial principal bundle $Q$ is locally weakly controllable near $q \in Q$ if and only if the space of Lie algebra $(\mathfrak{g})$ of structure group is spanned by the vector fields $\mathfrak{h}_1, \mathfrak{h}_2, \cdots$ as follows
\begin{equation}
\mathfrak{g}\:=\:\mathfrak{h}_1\:\oplus\:\mathfrak{h}_2\:\oplus \cdots 
\end{equation}

Whereas, the system is locally strongly controllable if and only if 
\begin{equation}
\mathfrak{g}\:=\:\mathfrak{h}_2\:\oplus\:\mathfrak{h}_3\:\oplus \cdots 
\end{equation}

The term corresponding to $\mathfrak{h}_1$ is just the space spanned by the columns of the local connection form $A(x)$, which in our case is a $3 \times 2$ matrix. $\mathfrak{h}_2$ is the column corresponding to the curvature $DA$, which is a differential 2-form over $T_xM$. The explicit calculation is done using $DA = dA - [A,A]$, where $d(*)$ is the exterior derivative, and $[*,*]$ is the Lie bracket of the columns of $A(x)$. The terms from $\mathfrak{h}_3$ onwards are the spaces spanned by the bracketing operation of terms from connection form and its curvature's, along with $L_X\xi$ term, which is a Lie derivative with respect to the vector is in the tangent of the shape space.  

For the Purcell's swimmer we explicitly calculated $A(x)$, followed by terms $\mathfrak{h}_1$, $\mathfrak{h}_2$, $\mathfrak{h}_3$ the Purcell's swimmer. The term $L_X\xi$ is evaluated using Cartan's magic formula [\cite{bloch2003nonholonomic}]. We found that the rank of $\mathfrak{h}_2\:\oplus\:\mathfrak{h}_3$ is always $3$, hence satisfying the strong controllability conditions at all the points. Since all of these terms have a large expressions and are unwieldy to mention in the paper, we did a numerical calculation at few points in shape space $M$ and have shown the results in the Appendix A.

\section{Symmetric Purcell's swimmer}
\begin{figure}[h!]
\centering
\includegraphics[scale=0.38]{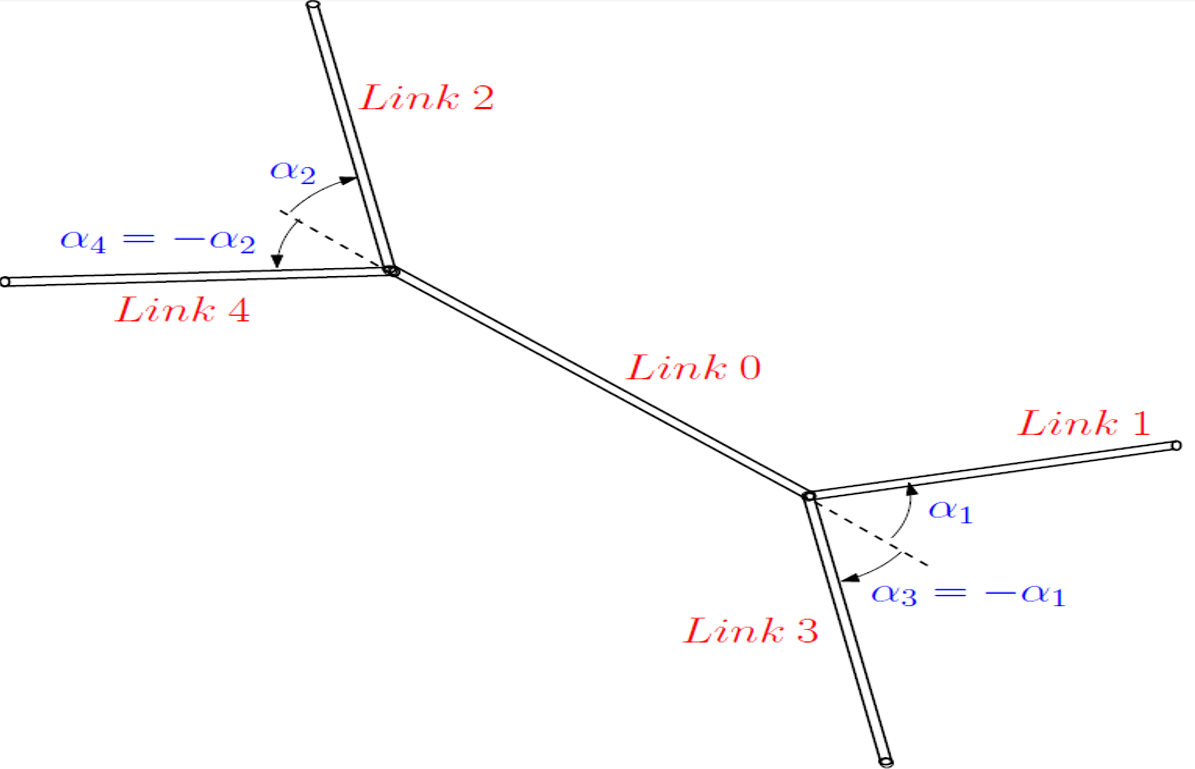}
\caption{Symmetric Purcell's swimmer}
\label{Sym_purcell}
\end{figure}

Now we turn our attention to the symmetrized version of the Purcell's swimmer, discussed in [\cite{avron2008geometric}]. Fig. \ref{Sym_purcell} shows the schematic of the swimmer. As compared to the original Purcell's swimmer, this has an additional limb at each of the 2 joints. Moreover, the limbs located at the same point rotate symmetrically about the base link $(\alpha_3 = - \alpha_1, \alpha_4 = -\alpha_2)$. Thus the shape space of this swimmer is 2 dimensional, parametrized by $(\alpha_1, \alpha_2)$. In the following section we derive the connection form for the symmetric Purcell's swimmer to show that it can perform motion in only one dimension, which is along the length of the base link. 

\subsection{Connection form}
In order to calculate the connection form, we extend the approach used for the basic Purcell's swimmer in [\cite{hatton2013geometric}]. Each of the limbs moves in the Special Eucledean group $SE(2)$. Hence we represent the velocity of i'th link by $\xi_i=[\xi_{i,x}, \xi_{i,y}, \xi_{i,\theta}]^T$. As a consequence of modelling of the limbs according to Cox theory [\cite{cox1970motion}] and the resistive force theory [\cite{tam2007optimal}] at low Reynold's number, for $k$ as the differential viscous drag, we get the force acting on each limb through linear relationship, 	

\begin{equation}
F_i=H \xi_i, \:H\:=\: \begin{bmatrix}
kL & 0 & 0 \\
0 & 2kL & 0 \\
0 & 0 & \frac{2}{3}kL^3 \\
\end{bmatrix}
\end{equation}

The velocity of the i'th link $\xi_i$ depends on the velocity of the base link and shape velocities $(\dot{\alpha}_1, \dot{\alpha}_2)$. This relationship is given by the following transformation -
\begin{equation}
\xi_i = B_i \begin{bmatrix}
\xi \\
\dot{\alpha}_1 \\
\dot{\alpha}_2
\end{bmatrix}, \:\:\:\: \xi = \begin{bmatrix}
\xi_{0,x} \\
\xi_{0,y} \\
\xi_{0,\theta}
\end{bmatrix}
\end{equation}
where,
\begin{align*}
B_1 &=
\begin{bmatrix}
\cos \alpha_1 & - \sin \alpha_1 & L\sin \alpha_1 & 0 & 0\\
\sin \alpha_1 & \cos \alpha_1 &  - L(\cos \alpha_1 + 1) & L & 0\\
0 & 0 & 1 & -1 & 0 
\end{bmatrix}, \\
B2 &=
\begin{bmatrix}
\cos \alpha_2 & \sin \alpha_2 & L\sin \alpha_2 & 0 & 0\\
-\sin \alpha_2 & \cos \alpha_2 &  - L(\cos \alpha_2 + 1) & L & 0\\
0 & 0 & 1 & 1 & 0 
\end{bmatrix}
\end{align*}

$B_3$ and $B_4$ are obtained by replacing $\alpha_1$ by $-\alpha_3$ and $\alpha_2$ by $-\alpha_4$ in $B_1$ and $B_2$ respectively, and multiplying the last 2 columns of $B_1$ and $B_2$ by $-1$, since the direction of motion of the symmetric limb would always be equal in magnitude and opposite in direction. Since the link 0 is the base link, the coordinate frame attached to which is the reference body frame, we get
\begin{equation}
B_0 =\begin{bmatrix}
1 & 0 & 0 & 0 & 0\\
0 & 1 & 0 & 0 & 0\\
0 & 0 & 1 & 0 & 0 
\end{bmatrix}
\end{equation}
We transform the force acting on the each link to the coordinate frame attached to the base link. The force transformation matrices for link 1 and 2 are given by
\begin{align*}
T_1&=\begin{bmatrix}
\cos \alpha_1 & \sin \alpha_1 & 0\\ 
- \sin \alpha_1 & \cos \alpha_1 & 0\\ 
L\sin \alpha_1 &  - L (\cos \alpha_1 + 1) & 1 
\end{bmatrix}, \\
T_2&=\begin{bmatrix}
\cos \alpha_2 & - \sin \alpha_2 & 0\\ 
\sin \alpha_2 & \cos \alpha_2 & 0\\ 
L\sin \alpha_2 & L(\cos \alpha_2 + 1) & 1
\end{bmatrix}
\end{align*}

Due to symmetry, these force transformation matrices for links 3 and 4 are obtained by replacing $\alpha_1$ by $-\alpha_3$ and $\alpha_2$ by $-\alpha_4$, respectively. Again, since the link 0 is the reference link, we get $T_0$ as the identity transformation. The total force is summation of the forces acting on the individual links, transformed to the coordinate frame of the base link.

\begin{equation}
F_{total} = \sum_{i=0}^4 T_i H B_i \begin{bmatrix}
\xi \\ 
\dot{\alpha}_1\\ 
\dot{\alpha}_2
\end{bmatrix}
\end{equation}

The consequence of being at low Reynolds number is that the net forces and moments on an isolated system is zero [\cite{hatton2013geometric}]. Moreover, in order to bring the system of equations in a pure kinematic form, [\cite{shammas2007geometric}], we write the terms $T_i H B_i$ in the block matrix form to separate those columns which are being multiplied by group velocity term $\xi$ and those being multiplied by shape velocity term $[\dot{\alpha}_1,\: \dot{\alpha}_2]^T$. This yields
\begin{equation}
0 = \sum_{i=0}^4 [[T_i H B_i]_{3\times3} \:\:\: [T_i H B_i]_{3\times2}] \begin{bmatrix}
\xi \\ 
\dot{\alpha}_1\\
\dot{\alpha}_2
\end{bmatrix}
\end{equation}
Thus we write the system in a pure kinematic form, like in equation \ref{pure_kinematic}, as 
\begin{equation}
\sum_{i=0}^4 [T_i H B_i]_{3\times3}
\xi = \sum_{i=0}^4 [T_i H B_i]_{2\times2} \begin{bmatrix}
\dot{\alpha}_1\\ 
\dot{\alpha}_2
\end{bmatrix}
\end{equation}
with local connection form as
\begin{equation}
A = [\sum_{i=0}^4 [T_i H B_i]_{3\times3}]^{-1} \sum_{i=0}^4 [T_i H B_i]_{3\times2}
\end{equation}

On explicit calculation, for $k =1, L =1$, we get connection form as a $2 \times 2$ matrix whose columns $\mathfrak{h}_1$ and $\mathfrak{h}_2$ are given as 
\begin{small}
\begin{equation}\label{columns_one_form}
\mathfrak{h}_1 = \begin{bmatrix}
\frac{4 \sin\alpha_1}{2 \sin^2 \alpha_1 + 2 \sin^2 \alpha_2 + 5}\\ 
0 \\ 
0 
\end{bmatrix},
\mathfrak{h}_2 = \begin{bmatrix}
\frac{-4 \sin\alpha_2}{2 \sin^2 \alpha_1 + 2 \sin^2 \alpha_2 + 5}\\
0 \\ 
0 
\end{bmatrix} 
\end{equation}

\end{small}

The rows of the connection form corresponding to velocity directions $\xi_y$ and $\xi_\theta$ are identically zero, which implies that the motion of the swimmer is always along the length of the base link. Thus the configuration space of the swimmer is obtained as $Q = \mathbb{R}^1 \times \mathbb{S}^1 \times \mathbb{S}^1$.

\subsection{Controllability of the symmetric Purcell's swimmer}
Since the structure group of the symmetric Purcell's swimmer is just the real line $\mathbb{R}$, which is an Abelian group, in order to prove the controllability we use the special case of the Ambrose-Singer theorem, referred from [\cite{kelly1995geometric}]. It says that a locomotion system on an Abelian principal bundle is strongly controllable at $x \in M$, if and only if
\begin{equation}
span\{dA(X,Y) : X,Y \in T_xM, x \in M \} = \mathfrak{g}
\end{equation}

A similar argument as that for the basic Purcell's swimmer shows that the symmetric Purcell's swimmer also follows a trivial principal fiber bundle structure. Hence, we can apply this test for its controllability. From equation (\ref{columns_one_form}), by taking the exterior derivative, we get the expression for $dA$, a differential 2-form as 
\begin{equation}
dA=\displaystyle \displaystyle \frac{16 \sin \alpha_1 \sin \alpha_2 (\cos \alpha_1 - \cos \alpha_2)}{ (2 \sin^2 \alpha_1 + 2 \sin^2 \alpha_2 + 5)^2} \:\: d\alpha_1 \wedge d\alpha_2
\end{equation}

We define following 3 sets
\begin{align*}
S1&=(\alpha_1, \alpha_2) \in M \setminus \{(0,*), (*,0), (\pi, *), (*, \pi), \\  
& \qquad  (c, c), \forall c \in (0,\pi)\} \\
S2&=(\alpha_1, \alpha_2) \in \{(0,0), \: (\pi,\pi)\} \\
S3&=(\alpha_1, \alpha_2) \in \{M \setminus S1 \} \setminus S2
\end{align*}

Thus we conclude the weak and strong controllability for the symmetric version of the swimmer as below
\begin{itemize}
\item  The system is strongly controllable when
\begin{equation}\nonumber
span \{dA\} = 1 \:\: \forall (\alpha_1, \alpha_2) \in S1
\end{equation}

\item  The system is not controllable when
\begin{equation}\nonumber
span\{\mathfrak{h}_1 \oplus \mathfrak{h}_2 \} = span\{dA\}= 0,  \forall (\alpha_1, \alpha_2) \in  S2
\end{equation}

\item The system is weakly controllable when
\begin{equation}\nonumber
span\{\mathfrak{h}_1 \oplus \mathfrak{h}_2 \} = 1, \:\: span\{dA\}= 0 \:\: \forall (\alpha_1, \alpha_2) \in  S3
\end{equation}

\end{itemize}
These 3 sets are represented in figure \ref{symm_purcell_set} below. 
\begin{figure}[h!]
\centering
\includegraphics[scale=0.4]{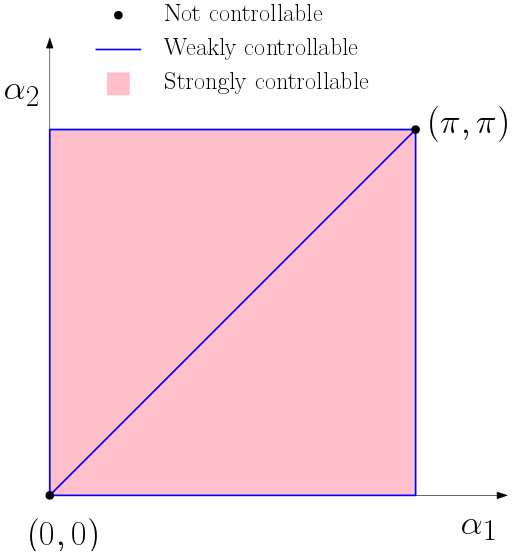}
\caption{Controllable sets of symmetric Purcell's swimmer}
\label{symm_purcell_set}
\end{figure}

\section{Conclusion and Future work}
In this paper we explored the geometry of the Purcell's swimmer's configuration and showed its strong controllability by using an extension of Chow's theorem for systems on a principal fiber bundle. We then used the resistive force theory to derive the kinematic form of the symmetric version of the Purcell's swimmer. We showed that its structure group is Abelian and used the Ambrose-Singer theorem to characterize the points in the configuration space at which the system is locally strongly and weakly controllable and uncontrollable.

This work and many other contributions in the Purcell's swimmers modelling, control and controllability revolve around the basic planar swimmer. Even the N-link extension of the Purcell's swimmer [\cite{giraldi2013controllability}] studies controllability and optimal gait design for planar swimming alone. It seems pertinent to explore the applicability of the underlying theory of low Reynold's number for slender members to 3-dimensional motion. This extension can open up many avenues for replicating more realistic microbial motion, followed by subsequent mechanistic and control theoretic analysis.

\bibliography{ifacconf}
\appendix
\section{Basic Purcell's Controllability condition}
We take a few points from the configuration space of the basic Purcell's swimmer, and tabulate numerical values of the terms $\mathfrak{h}_1$, $\mathfrak{h}_2$, $\mathfrak{h}_3$, explained in section 2. It is to be noted that differential viscous drag $k$ and half of the limb lengths $L$ are both taken to be unity for simplicity; the controllability results still hold the same for any other realistic value. We note that the dimension of the space spanned by these terms is $3$ at all these points.

\begin{small}
\begin{tabular}{m{0.5cm} m{4cm} m{2.5cm} m{4.5cm} m{2.5cm}}
\toprule
\multicolumn{1}{c}{$(\alpha_1, \alpha_2)$} & 
\multicolumn{1}{c}{$ \mathfrak{h}_1$} & 
\multicolumn{1}{c}{$\mathfrak{h}_2$} & 
\multicolumn{1}{c}{$\mathfrak{h}_3$} & 
\multicolumn{1}{c}{$span\{\mathfrak{h}_2\oplus\mathfrak{h}_3\}$} \\ 
\midrule
$(0,0)$ &
\adjustbox{valign=c}{%
\{$\begin{bmatrix} 
0	\\
0.333	\\
-0.259\\
\end{bmatrix}$
, $\begin{bmatrix}
0	\\
0.333	\\
0.259
\end{bmatrix}$
\}} &
\adjustbox{valign=c}{
\{$\begin{bmatrix} 
-0.469	\\
0	\\
0\end{bmatrix}$
}\}& \hspace{11pt}
\adjustbox{valign=c}{%
\{$\begin{bmatrix} 
0	\\
0.228	\\
-0.242\end{bmatrix}$
, %
  $\begin{bmatrix} 
0	\\
0.821	\\
-0.329\end{bmatrix}$
\}} & \hspace{30pt}
$3$\\

\vspace{10pt}
$(0,\pi/4)$ &
\vspace{10pt}
\adjustbox{valign=c}{%
\{$\begin{bmatrix} 
-0.106	\\
0.317	\\
-0.277	\end{bmatrix}$
, $\begin{bmatrix}
-0.226	\\
0.306	\\
0.243\end{bmatrix}$
\}} &
\vspace{10pt}
\adjustbox{valign=c}{
\{$\begin{bmatrix} 
-0.389	\\
-0.185	\\
-0.058\end{bmatrix}$
}\}& \vspace{10pt}
\hspace{11pt}
\adjustbox{valign=c}{%
\{$\begin{bmatrix} 
-0.087	\\
0.233	\\
-0.237\end{bmatrix}$
, %
  $\begin{bmatrix} 
-0.341	\\
0.778	\\
-0.348\end{bmatrix}$
\}} & 
\vspace{10pt}
\hspace{30pt}
$3$\\

\vspace{10pt}
$(0,\pi/2)$ &
\vspace{10pt}
\adjustbox{valign=c}{%
\{$\begin{bmatrix}
-0.16	\\
0.271	\\
-0.320	\end{bmatrix}$
, $\begin{bmatrix}
-0.395	\\
0.283	\\
0.209	\\
\end{bmatrix}$
\}} &
\vspace{10pt}
\adjustbox{valign=c}{
\{$\begin{bmatrix} 
-0.223	\\
-0.321	\\
-0.064	\end{bmatrix}$
}\}& \vspace{10pt}
\hspace{11pt}
\adjustbox{valign=c}{%
 \{$\begin{bmatrix} 
-0.024	\\
0.308	\\
-0.167\end{bmatrix}$
, %
  $\begin{bmatrix} 
-0.339	\\
0.812	\\
-0.301\end{bmatrix}$
\}} & 
\vspace{10pt}
\hspace{30pt}
$3$\\

\vspace{10pt}
$(0,3 \pi/4)$ &
\vspace{10pt}
\adjustbox{valign=c}{%
\{$\begin{bmatrix} 
-0.127	\\
0.222	\\
-0.366
\end{bmatrix}$
, $\begin{bmatrix}
-0.416	\\
0.714	\\
0.343
\end{bmatrix}$
\}} &
\vspace{10pt}
\adjustbox{valign=c}{
\{$\begin{bmatrix} 
-0.289	\\
-0.317	\\
0.037	\end{bmatrix}$
}\}& \vspace{10pt}
\hspace{11pt}
\adjustbox{valign=c}{%
\{$\begin{bmatrix} 
-0.033	\\
1.261	\\
0.227\end{bmatrix}$
, %
  $\begin{bmatrix} 
-0.418	\\
2.828	\\
0.325\end{bmatrix}$
\}} & 
\vspace{10pt}
\hspace{30pt}
$3$\\

%\vspace{10pt}
$(0,\pi)$ &
\vspace{10pt}
\adjustbox{valign=c}{%
\{$\begin{bmatrix} 
0	\\
0.333	\\
-0.333	\end{bmatrix}$
, $\begin{bmatrix}
0	\\
5.66	\\
2.33	\end{bmatrix}$
\}} &
\vspace{10pt}
\adjustbox{valign=c}{
\{$\begin{bmatrix} 
-2.888	\\
0	\\
0	
\end{bmatrix}$
}\}& \vspace{10pt}
\hspace{11pt}
\adjustbox{valign=c}{%
\{$\begin{bmatrix} 
0	\\
34.481	\\
12.481
\end{bmatrix}$
}, {%
  $\begin{bmatrix} 
0	\\
77.629	\\
10.518
\end{bmatrix}$
\}} & 
\vspace{10pt}
\hspace{30pt}
$3$\\

\vspace{10pt}
$(\pi,0)$ &
\vspace{10pt}
\adjustbox{valign=c}{%
\{$\begin{bmatrix} 
0	\\
-0.629	\\
0.185
\end{bmatrix}$
, $\begin{bmatrix}
0 \\
-0.037 \\
0.481\end{bmatrix}$
\}} &
\vspace{10pt}
\adjustbox{valign=c}{
\{$\begin{bmatrix} 
0.419	\\
0	\\
0	\end{bmatrix}$
}\}& \vspace{10pt}
\hspace{11pt}
\adjustbox{valign=c}{%
\{$\begin{bmatrix} 
0	\\
-0.059	\\
-0.372\end{bmatrix}$
}, {%
  $\begin{bmatrix} 
0	\\
-0.166	\\
-0.232\end{bmatrix}$
\}} & 
\vspace{10pt}
\hspace{30pt}
$3$\\
\bottomrule 
\end{tabular}
\end{small}
\end{document}